\documentclass[referee]{aa}

\usepackage[varg]{txfonts}
\usepackage{lineno}
\usepackage{natbib,twoopt}
\usepackage[breaklinks=true]{hyperref} %% to avoid \citeads line fills
\bibpunct{(}{)}{;}{a}{}{,} %% natbib format for A&A and ApJ
\makeatletter
\newcommandtwoopt{\citeads}[3][][]{\href{http://adsabs.harvard.edu/abs/#3}%
{\def\hyper@linkstart##1##2{}%
\let\hyper@linkend\@empty\citealp[#1][#2]{#3}}}
\newcommandtwoopt{\citepads}[3][][]{\href{http://adsabs.harvard.edu/abs/#3}%
{\def\hyper@linkstart##1##2{}%
\let\hyper@linkend\@empty\citep[#1][#2]{#3}}}
\newcommandtwoopt{\citetads}[3][][]{\href{http://adsabs.harvard.edu/abs/#3}%
{\def\hyper@linkstart##1##2{}%
\let\hyper@linkend\@empty\citet[#1][#2]{#3}}}
\newcommandtwoopt{\citeyearads}[3][][]%
{\href{http://adsabs.harvard.edu/abs/#3}
{\def\hyper@linkstart##1##2{}%
\let\hyper@linkend\@empty\citeyear[#1][#2]{#3}}}
\makeatother

\begin{document}

\title{Space, time and velocity association of  successive coronal mass ejections}

\titlerunning{Dependence of Consecutive CMEs}

\author{
Alejandro Lara \inst{1,2}
\and Nat Gopalswamy \inst{3}
\and Tatiana Niembro \inst{4}
\and Rom\'an P\'erez-Enr\'iquez \inst{5}
\and Seiji Yashiro \inst{2} }

\institute{Instituto de Geof\'isica, Universidad Nacional Aut\'onoma de M\'exico,  Ciudad de MÃ©xico,  M\'exico \and The Catholic University of America, Washington, DC, USA \and GSFC/NASA, Greenbelt MD, USA \and Smithsonian Astrophysical Observatory, Cambridge, MA, USA \and Centro de Geociencias, Universidad Nacional Aut\'onoma de M\'exico,  Quer\'etaro, M\'exico
}

%\date{Received 2 November 1992 / Accepted 7 January 1993}

\abstract
{}
{Our aim is to investigate the possible physical association between
 consecutive  coronal mass ejections (CMEs).}
{Through a statistical study of the main characteristics of  27761 CMEs observed by SOHO/LASCO during the past 20 years.}
  {We found the waiting time (WT) or time elapsed between two consecutive CMEs is $< 5$  hrs for 59\% and $< 25$ hrs for 97\% of the events, and the CME WTs follow a Pareto Type IV statistical distribution. The difference of the position-angle of a considerable population of consecutive CME pairs is less than $30^\circ$, indicating the possibility that their source locations are in the same region. The difference between the speed of trailing and leading consecutive CMEs follows a generalized Student t-distribution. The fact that the WT and the speed difference have heavy-tailed distributions along with a detrended fluctuation analysis shows that the CME process has a long-range dependence. As a consequence of the long-range dependence, we found a small but significative difference between the speed of consecutive CMEs, with the speed of the trailing CME being higher than the speed of the leading CME. The difference is largest for WTs < 2 hrs and  tends to be zero for WTs > 10 hrs, and it is more evident during the ascending and descending phases of the solar cycle.
We suggest that this difference may be caused by a drag force acting over CMEs
closely related in space and time.}
{Our results show that the initiation and early propagation  of  a significant population of CMEs 
 cannot be considered as a 
``pure'' stochastic process; instead they have temporal, spatial, and velocity relationship.}


\maketitle

\section{Introduction}
\label{sec:intro}

Coronal mass ejections (CMEs) were discovered in the 1970s by the observations of the white-light coronograph experiment on board  the OSO-7 mission \citep{1973SoPh...33..265T}. Since then, a large number of 
data have been accumulated that help us understand the basic characteristics of CMEs, but many questions still remain unanswered \citep[see e. g.,][and references therein]{2004ASSL..317..201G,2012LRSP....9....3W,2016GSL.....3....8G}. In particular, the Large Angle and Spectrometric Coronograph \citep[LASCO,][]{1995SoPh..162..357B} on board the Solar and Heliospheric Observatory (SOHO) spacecraft  has  been  very successful in contributing considerably to this understanding of CME characteristics. In this work, we use the high number of LASCO observations and the CME characteristics, measured and made available online in the CDAW database \citep{2009EM&P..104..295G}, to explore the statistical properties of CMEs. In particular, we are interested in consecutive CMEs, which may help in the understanding of the energy storage and its release via a triggering mechanism,  and the  early stages of the CME dynamics.
Coronal mass ejections
and flares consume only part of the free magnetic energy available in active regions \citep{2013ApJ...765...37F}. The released energy (considering the flare and CME energy) may be, on average, a third of the free magnetic energy \citep{2012ApJ...759...71E}. Furthermore, CMEs carry  on average 7\%  of the total dissipated energy during the eruptive event  \citep{2017ApJ...836...17A}. Therefore, in terms of the stored energy, more than one CME can be ejected from the same active region \citep{2005JGRA..110.9S15G,2013AdSpR..52..521M,2017arXiv170903165G}, and more importantly,  within  short waiting times (WT), which is the time elapsed between the launch of two successive CMEs (see Section \ref{sec:wt}).
The triggering mechanisms remain unsolved \citep{2000JGR...10523153F}. Recently, it has been proposed that CMEs may destabilize nearby active regions and trigger  ``sympathetic'' CMEs \citep[see][and references therein]{2017SoPh..292...64L}, in such cases short WT are expected, although, to support this causality, the WT distribution should be different from one of a Poisson random process.
Therefore, the WT distribution  gives valuable information about the so-called ``memory'' of the system. For memory-less systems (stochastic process) the distribution is exponential or gamma. On the other hand, the so called ``fat-tailed'' distributions (e.g., Power Law, Pareto, Weibull, etc) are often associated with memory systems \citep[see][for a general discussion of this subject]{2005Natur.435..207B,Samorodnitsky2007,SHENG2011}.
As evidenced by coronal EUV dimmings, the footprints of CMEs  \citep{2000GeoRL..27.1431T}, the corona behind CMEs is ``evacuated'' \citep{2017ApJ...839...50K} during  a few hours
\citep{2017SoPh..292....6L}. Therefore,  a second CME launched in the same region within a short WT encounters a different coronal medium than the first CME, and its dynamics are different according to  the drag force models of CME propagation \citep{2004SoPh..221..135C,2009A&A...498..885B,2010A&A...512A..43V}.
If  the second CME is faster than the first one, and the source region is similar, there is a high probability of interaction at any heliospheric distance. This interaction  causes CME cannibalism \citep{2001ApJ...548L..91G}, and the merged regions have a  different dynamics than a single CME \citep{2015ApJ...811...69N}, making the prediction of the travel time and arrival speed at 1 AU difficult \citep{2013SpWea..11..661G}.
The paper is organized as follows: The SOHO/LASCO data used in this work is presented in Section \ref{sec:db}. We determine the waiting time (WT) between consecutive CMEs, and present statistical characteristics in Section \ref{sec:wt}. The WT has a strong dependence  on the solar cycle phase,
which we describe in Section \ref{sec:phases}.
We study  the source region of consecutive CMEs, analyzing the statistical behavior of the difference of position angle (PA)  (Section \ref{sec:pa})  and speed (Section \ref{sec:speed})  between consecutive CME pairs.
Given that both WT and speed have heavy-tailed distributions, which are associated with long-range dependence, we perform a detrended fluctuation analysis to corroborate the long-range dependence in Section \ref{sec:dfa}.
This dependence is clearly seen in the difference between the speed of trailing and leading CMEs (Section \ref{sec:speed_rel}).
Finally, our discussion and conclusions are presented in
Sections \ref{sec:disc} and \ref{sec:conclusions}, respectively.

\section{CME database}
\label{sec:db}

The CME database is maintained by the CDAW team at the CDAW Data Center, NASA Goddard Space Flight Center (https://cdaw.gsfc.nasa.gov). The database includes the basic sky-plane measurements of CMEs such as speed, angular width, central position angle, acceleration, mass, and kinetic energy. The first-appearance time of each CME in the LASCO field of view (FOV) is also given in the catalog, which we use for determining the WT. The observations have been continuously available since 1996, with a major three-month gap in 1998 when the SOHO spacecraft was temporarily disabled \citep{2004JGRA..109.7105Y,2008AnGeo..26.3103Y,2009EM&P..104..295G}.
 In order to avoid possible instrumental errors during the first stage of the LASCO mission, from the 29577 CMEs recorded from January 1996 until the end of December 2018, in this work we consider only 28270 CMEs observed after the second major data gap, starting with the CME observed on February 2, 1999 at 16:35, and 
finishing with the CME observed on December 31, 2018 at 09:48 UT.
Taking into account that our analysis is performed over pairs of consecutive CMEs, using the second CME of each pair as a reference, the number of pairs is equal to the number of CMEs minus one.
Furthermore, we skip the first CME after a data gap (which  corresponds to the
second CME of the registered pair during the data gap), this way discarding the data gaps from the analysis 
(it is important to note that we do not apply any restriction to select the leading-trailing CME pairs, so that a leading one can be a trailing one of a previous event, and these are treated as an independent pair).
During the entire period (1996-2018),  1003 data gaps longer than two hours were reported, from these data gaps, 673 occurred during our period of study.
 Although, in some cases (specifically, 164), more than one data gap took place in between two observed CMEs.
Therefore, our final set contains 27761 CMEs.

\section{Waiting time}
\label{sec:wt}

\begin{figure}
\includegraphics[width=\columnwidth]{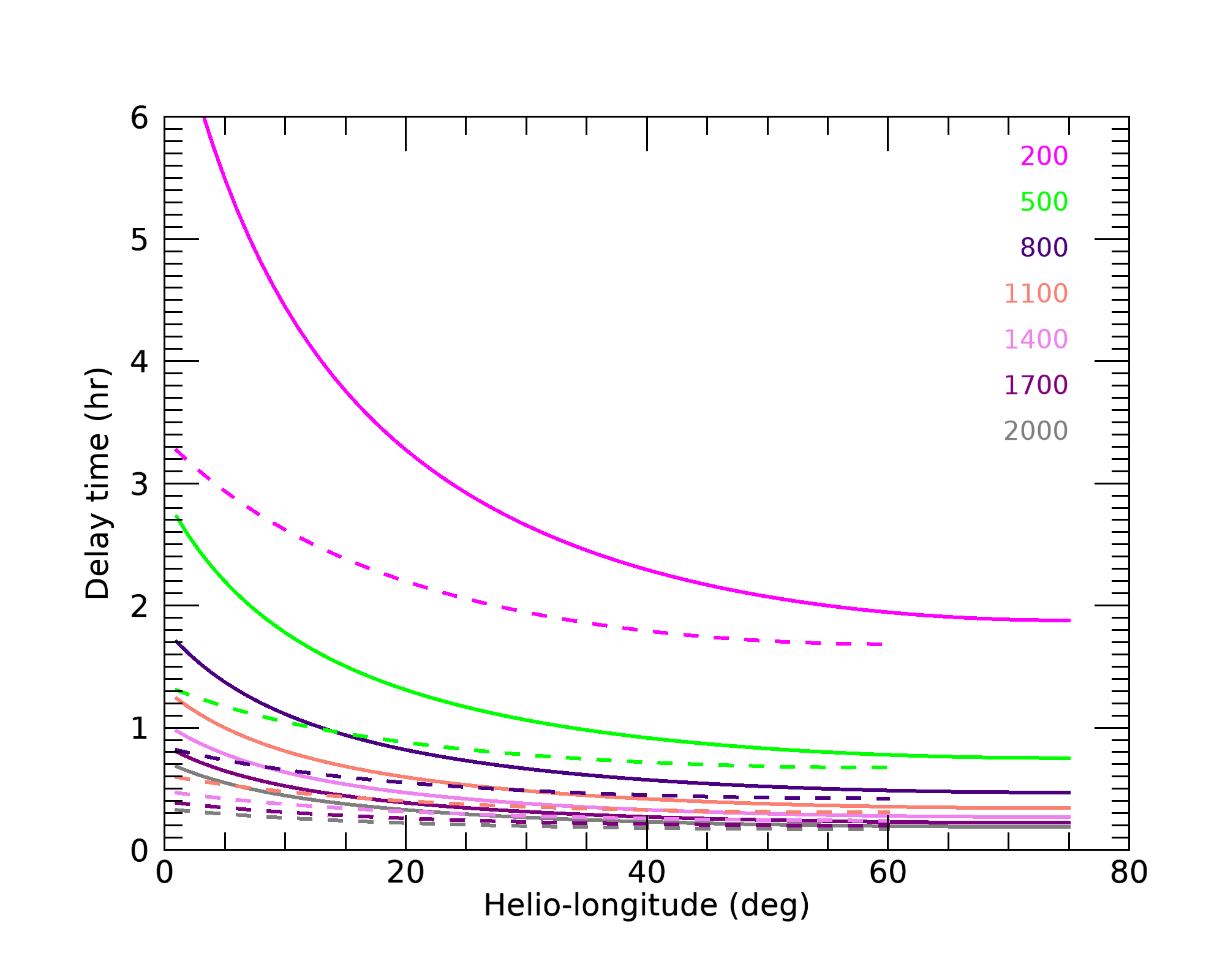}
\caption{Delayed time of first observation of a CME with constant speed, occulted by a 2 $R_\odot$ disk as a function of the helio-longitude of the source region. Continuous and dashed lines correspond to CME width of $30^\circ$ and $60^\circ$, respectively.}
\label{fig:tdelay}
\end{figure}

We define the CME WT as the time elapsed between the first observation of two consecutive CMEs in the LASCO FOV. It is important to note that the CME initial time, as reported when using coronographic observations, is not exactly the launching time, but the time when a CME first appears in the coronograph FOV above the occulting disk. Therefore, there is a small non-uniform error in assigning the initial time.  This error is smaller for limb CMEs and increases when the source region of the CME approaches the center of the disk. This effect can be seen in Figure \ref{fig:tdelay}, where we show the delay time of a CME observed by a coronograph with a two solar radii ($R_\odot$) occulting disk. This is equivalent to the time that it takes for a CME with given constant speed (from 200 to 2000 km s$^{-1}$, marked by colors in the figure) to travel a distance of 2 $R_\odot$ projected in the plane of the sky,  as a function of the helio-longitude of the CME source region. The delay time ($t$) is given as
 $t= \left[ \sin(\theta) + \tan(w/2) \cos(\theta) \right]^{1/2} R_\odot/v$,
 where $\theta$ is the helio-longitude, and $w$ and $v$ are the CME width and speed, respectively. 
 This analysis shows that for very slow CMEs ($V \leq 200$ km s$^{-1}$), the delay time is $< 1$ hr when the source region is located beyond $45^\circ$ of helio-longitude, whereas for very fast CMEs ($V > 1000$ km s$^{-1}$), the delay time may be considered constant for helio-longitudes higher than $30^\circ$. We note that this basic analysis assumes similar visibility for front- and back-sided CMEs.
The delay time is also shorter for wider CMEs for a given speed.

As shown in Figure \ref{fig:tdelay}, narrow and slow CMEs launched close to the center of the disk (lon $\ge \pm30^\circ$) have a delay of the observed initial time longer than two hours. Although, CMEs may measure any width and may be launched in any longitude (the center/limb launching  position is an observational effect, and thus, the probability that a CME of a given speed is launched inside this longitude range  is $\sim 1/3$). Therefore, as we have a large number of events (launched with different velocities, widths and longitudes), by selecting a minimum of two hours for the WT, we are able to neglect the error introduced by the slow-narrow CMEs launched close to the center of the disk.

Another source of systematic errors is the cadence of the instrument. In the case of LASCO, this is between 12 and 40 minutes (depending on the observation mode). Therefore, WTs $< 1$ hr are difficult to quantify properly. On the other hand, only  slow ($V < 600$ km s$^{-1}$) and central source region ($|\theta| < 15^\circ$) CMEs have delays $ >  2$ hrs (Figure \ref{fig:tdelay}). Therefore, for the majority of CMEs the delay is under two hours, and this can be taken as the WT resolution. As a consequence, the bin size of the distributions in the next section has been taken as two hours.
We note that the cadence of LASCO was doubled  after August 2010, this caused a steep elevation of the CME rate reported in automated (CACTus and SEEDS), but not in CDAW catalogs \citep{2014ApJ...784L..27W,2017ApJ...836..134H}. This is aside from the fact that the real number of CMEs was higher for cycle 24 than for cycle 23 \citep{2015ApJ...812...74P}. The cadence change affected only very narrow CMEs, while wide  CMEs such as halos were not affected \citep{2013SpWea..11..661G}.

%\subsection{Waiting time distribution}

The statistical  distributions are useful for understanding the physical nature of the underlying process. In particular, the CME WT may help to determine whether the CME process is purely stochastic or  if there is a dependence or  physical connection between consecutive CMEs.
This is an important question in terms of CME triggering, and therefore a major issue in terms of prediction of space weather.

\begin{figure} 
\includegraphics[width=\columnwidth]{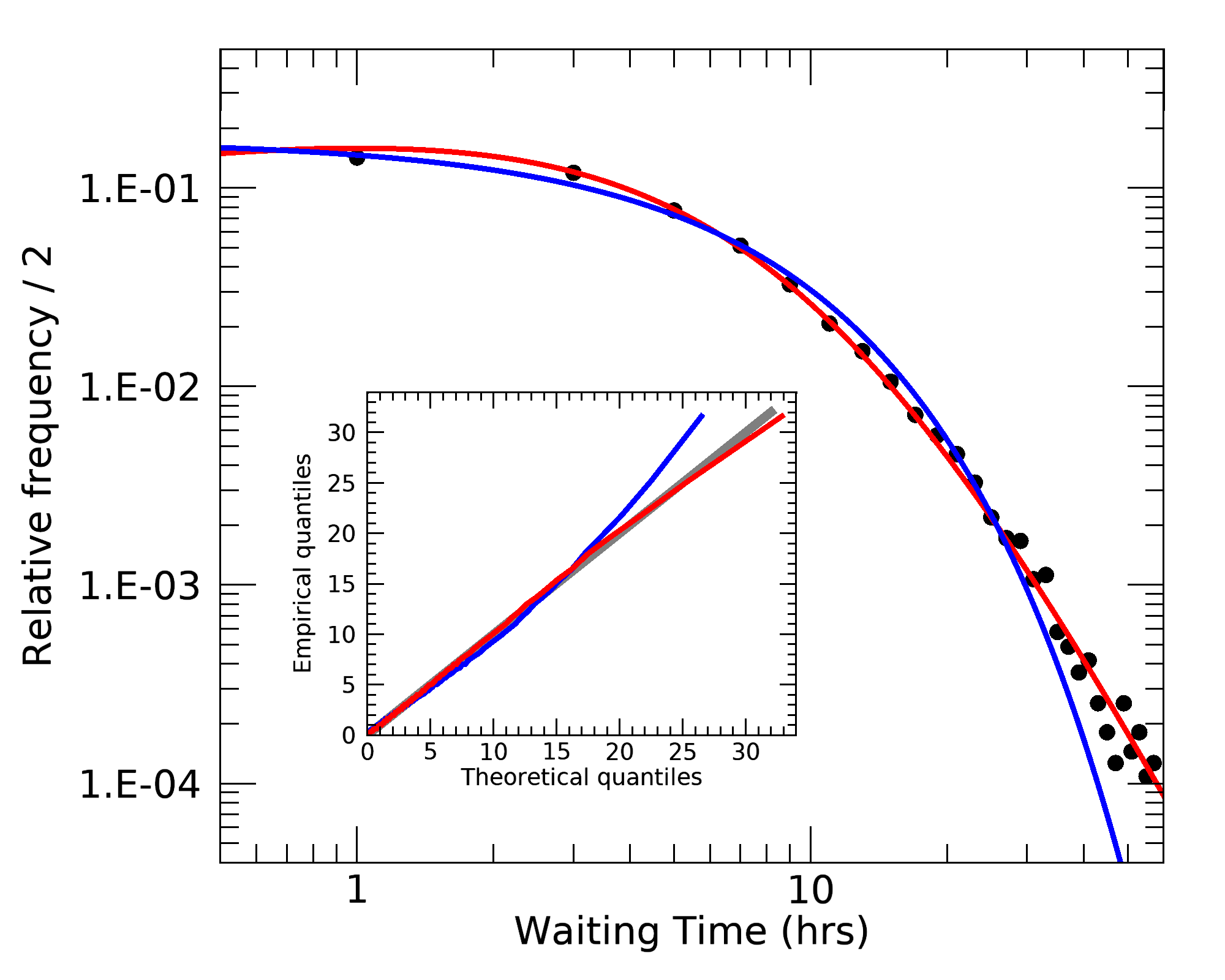}
\caption{%Normalized
Relative frequency (divided by the bin size of 2 hrs) of observed WT distribution (black circles), along the fit  exponential (blue) %, Gamma (cyan),
and Pareto (red) distributions. The inner plot is the so-called ``Q-Q plot,'' which graphically  illustrates that the  Pareto distribution (red line) is closer to the ``true'' distribution (represented by the gray line) than the exponential distribution.}
\label{fig:rep-rate}
\end{figure}

 Figure \ref{fig:rep-rate} shows the distribution of the WT (black
 circles) of
 27761 CMEs observed by LASCO from 
 February 1999 up to December 2018.
%
%%%%%
%
The main characteristics of the WT distribution are:
i) Almost all CMEs (98\%) occurred inside a time interval of 25 hours. Only  567  CMEs have a time difference longer than 25 hours.
ii) $\sim$  85\% of the events have a time difference shorter than 10 hours.
iii) The WT is under five hours for $\sim 61\%$ of the events.

Using a maximum likelihood method, we fit an exponential and a Pareto Type IV distribution to the observed WT. These distributions are plotted with  blue and red continuous lines, respectively,
in Figure \ref{fig:rep-rate}. Their probability density functions (PDF) and best fit parameters found through a maximum likelihood method are: 
\begin{itemize}

\item Exponential distribution  (blue curve):
\begin{equation}
  f(x)=\lambda \exp^{-x \lambda}   ,  \label{eq:expdist} 
\end{equation} 
with $\lambda = 0.17 $
 and a mean value of   5.75.

\item
Pareto Type-IV distribution  (red curve):
\begin{equation}
f(x)= \frac{ \kappa^{-1/\gamma} \alpha}{\gamma } \left[ 1 + \left( \frac{\kappa}{
    x - \mu} \right)^{-1/\gamma}\right]^{-1 - \alpha}  (x - \mu)^{-1 + 
  1/\gamma}  ,   \label{eq:paretodist}
\end{equation} 
with
$\kappa= 10.49$, $\alpha=2.84$  , $\gamma=0.81,$  and $\mu=0.07$,
 and a mean value of
5.79.
\end{itemize}

The Exponential distribution is memory-less and arises from a ``simple'' stochastic process like the waiting times of a Poisson process \citep{marshall2007life};
the memory-less property implies that there is no dependence or precondition between consecutive events. On the other hand, the Pareto distribution has a heavy-tail\footnote{Heavy-tailed distributions are characterized by the slow decreasing tail as compared with the exponential distribution}  \citep{marshall2007life} and has been associated with long-range dependence or long-memory  processes in a large variety of systems \citep{Samorodnitsky2007}.
It is a very interesting property  often associated with nonstationary processes,  and the scaling and fractal behavior of the system and  phase transitions \citep[see][for a survey of different points of view of the long-range dependence]{Samorodnitsky2007}.
As shown in Figure \ref{fig:rep-rate}, the Pareto distribution (red line) better follows the observed WT (black circles) in the entire range. This  fact is better demonstrated by the quantile-quantile or ``Q-Q plot'' on the inner frame of Figure \ref{fig:rep-rate}.
Q-Q plots are used to graphically assess the quality of the fit between the model and  the empirical distribution, which is represented as a straight gray line in this case.
To quantify the differences between the proposed and the observed distributions, we use the Bayesian information criterion \citep[BIC, ][]{1978Schwarz}, which uses the maximum likelihood to determine the deviation between the empirical and the proposed distributions, with a penalty term for the number of parameters of the distributions.    
The distribution favored by BIC ideally corresponds to the candidate model which is a posteriori
 Therefore, the relevant indicator of the BIC is the relative change. The fitting results are shown in the last row of Table \ref{table:wt-dist}, where the  Pareto - exponential $\Delta$BIC is shown in the last column. The fit parameters of the Pareto Type IV and exponential are shown in columns 2-5 and 7, respectively.
As expected and  clearly seen in Figure \ref{fig:rep-rate-ssn}, the WT changes appreciably during the different phases of the solar cycle. This causes the nonstationary aspect of the process. Therefore, an analysis taking into account the changes on each phase of the solar cycle is necessary.

\section{Waiting time during the solar cycle}
\label{sec:phases}

\begin{figure}
\includegraphics[width=\columnwidth]{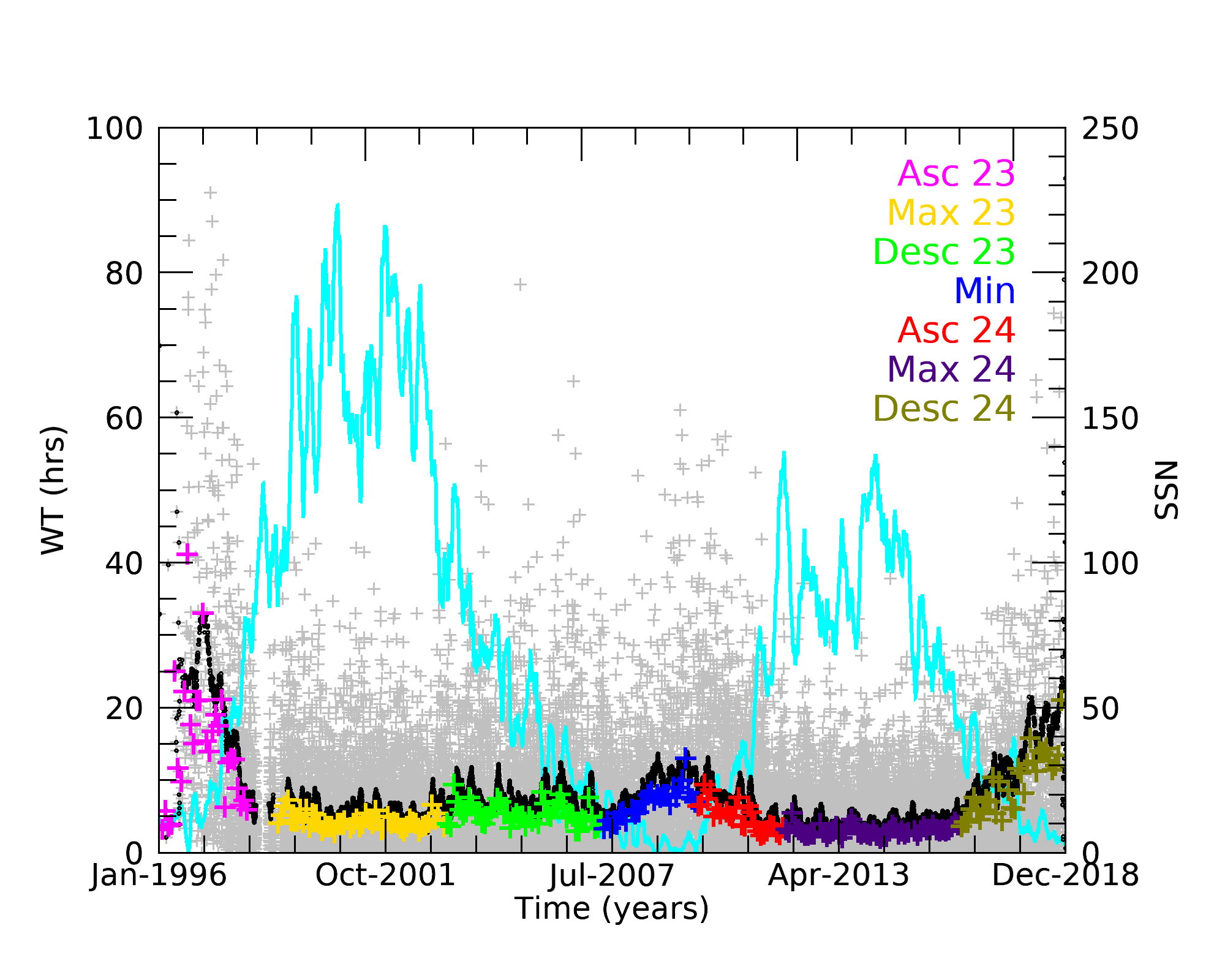}
\caption{WT between every two consecutive CMEs (gray plus symbols) as a function of time during the analyzed period.  The black dots correspond to a smoothed running average of the WT, whereas the cyan line corresponds to the Sunspot number. The different phases of the solar cycle (upper right) are marked with colored plus symbols, which correspond to the 30-day WT average.} 
\label{fig:rep-rate-ssn}
\end{figure}

\begin{figure}
\includegraphics[width=\columnwidth]{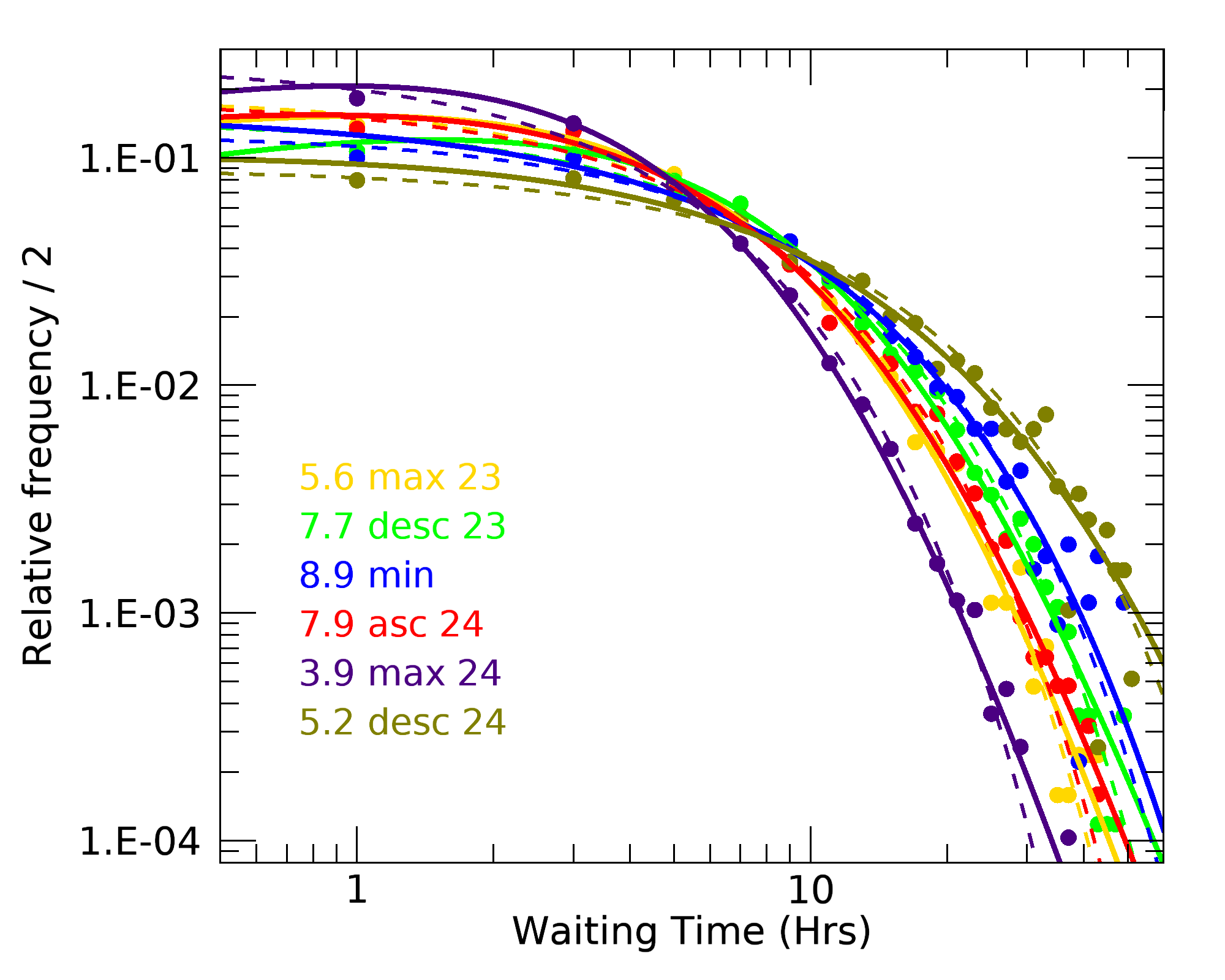}
\caption{%Statistical
Similar to Figure \ref{fig:rep-rate} but for 
different phases of solar cycles 23 and 24,
marked by
colors. The continuous and dash lines correspond to Pareto Type IV and
exponential distributions, respectively.
}
\label{fig:rr-cycle}
\end{figure}

The WT varies with time and follows the solar cycle, as seen in Figure \ref{fig:rep-rate-ssn}, where we have plotted the CME WT as a function of time (gray plus symbols); its smoothed version (running average of 60 points,  black dots) and a 30-day WT average (colored plus symbols), which shows a clear differentiation of the WT during the different solar cycle phases.
In order to compare these changes with other solar cycle parameters, we plotted the sunspot number with a cyan line (taken from WDC-SILSO, Royal Observatory of Belgium, Brussels). In general, the shortest WTs correspond to periods of maximum solar activity and viceversa. In fact, during the maximum of activity of  solar cycle 24, the mean WT was $ \leq 4$  hrs, whereas the mean WT reaches  $\sim 11$  hrs during the descending phase of cycle 24.
There are clear differences of the CME WT rate evolution during the solar cycle,  the 30-day average of the WT shows major changes along the time during low activity phases, whereas for high activity phases, it remains relatively constant. Following these changes, we divided the observed data in seven periods as marked in the upper right of Figure \ref{fig:rep-rate-ssn} with different colors.
The ascending phase of solar cycle 23 corresponds to the first stage
of LASCO, is somehow less stable than the other phases (as seen by the magenta symbols in Figure  \ref{fig:rep-rate-ssn}) and also contains the two major LASCO data gaps. Therefore, we did not take into account this phase for the rest of the analysis.
The observed WT distributions for each selected phase of the cycle are plotted in Figure \ref{fig:rr-cycle} with colored circles. Similarly to the entire time range WT distribution, we fit Pareto Type IV (Eq. \ref{eq:paretodist}) and an exponential (Eq. \ref{eq:expdist}) distributions.

There are important differences between the Pareto Type IV distribution parameters at each phase of the solar cycle. The wider WT distribution  (mean = 11.23 hrs) is associated with the descending phase of cycle 24.
The minimum phase of the cycle also has a wide  WT distribution (mean = 7.88 hrs). On the other hand, the maximum of 24 has the narrowest distribution (with a mean of 3.90 hrs). This is in accordance with the increase of the number of CMEs observed in cycle 24 \citep{2015ApJ...812...74P}. 
Table \ref{table:wt-dist} shows the parameters of the Pareto Type IV (columns 2 to 5) and exponential (column 7) distributions that best fit the WT during each phase of the solar cycles.

\begin{table*}
    \centering
      \caption{WT Distribution Parameters} \label{table:wt-dist}
\begin{tabular}{r|rrrrr|rrr}
  & \multicolumn{5}{c}{Pareto Type-IV}  &
                                             \multicolumn{3}{c}{Exponential}\\
  \hline
   Phase & $\kappa$ & $\alpha$ & $\gamma$ &  $\mu$ & mean & $\lambda$
                                                   & mean & $\Delta$  BIC \\
 \hline
%Asc 23   & 30.35 & 3.13 & 0.92 & 0.14 & 14.55 & 0.0691877 & 14.45 & -54.87 \\ 
Max 23  & 18.87 & 5.11 & 0.84 & 0.07 & 5.43 & 0.184542 & 5.42  & -259.40 \\ 
Desc 23 & 18.13 & 4.00 & 0.80 & 0.10 & 6.91 & 0.144912 & 6.90 & -218.20 \\ 
Min        & 182.73 & 23.34 & 1.02 & 0.20 & 7.88  & 0.126992 & 7.87 & -104.30 \\ 
Asc 24  & 19.94 & 5.03 & 0.87 & 0.07 & 5.65  & 0.177439 & 5.64  & -99.20 \\ 
Max 24 & 11.46 & 4.52 & 0.79 & 0.07 & 3.90 & 0.256769 & 3.89 & -632.80 \\ 
Desc 24 & 68.34 & 7.25 & 0.99 & 0.17 & 11.23 & 0.0890623 & 11.23 & -73.80 \\ 
\hline
 P3-P8  & 10.49 & 2.84 & 0.81 & 0.07 & 5.79 & 0.173847 & 5.75 & -1727.00 \\ 
%Total    & 9.41 & 2.49 & 0.81 & 0.07 & 6.10  & 0.165144 & 6.06 & -2232.00 \\ 
\end{tabular}
\end{table*}

The BIC shows that the observed WT distributions follow better the Pareto Type IV distribution. The differences between these distributions are  smaller during the phases of low activity (minimum and descending phase of cycle 24),
pointing towards a simple stochastic WT process. On the other hand, during high activity periods, 
the BIC differences clearly show that the WT follows the heavy-tailed Pareto Type IV distribution pointing towards a long-range dependence or memory process.
This long-range dependence of the WT time  maybe reflected in other characteristics of the CME phenomena. Therefore, we explore the source region and speed of consecutive CMEs in the following sections.

\section{Angular (PA) difference of consecutive CMEs}
\label{sec:pa}

The CME source region  may be used to explore the possible relationship between consecutive CMEs. In this analysis, we use the CME position angle (PA) as a proxy of the CME source region \citep{2008ApJ...688..647L}. In particular, we use the central position angle (CPA), which is the mid-angle between the CME edges\footnote{The CDAW catalog also includes the  measurement position angle (MPA), where the height-time measurements were taken \citep {2004JGRA..109.7105Y}: we do not use this angle in the present study.}. Out of the total number of observed CMEs, we excluded halo CMEs (which, by definition, do not have CPAs) and performed the present analysis over  26086 CMEs.
The rationale is that if there is long-range dependence in the CME WT, it should be statistically reflected in the distribution of the angular difference between the PA of consecutive CMEs ($PA_{diff}$). If the source regions of both consecutive events are relatively close, then the $PA_{diff}$ must approach zero.

\begin{figure}
\includegraphics[width=\columnwidth]{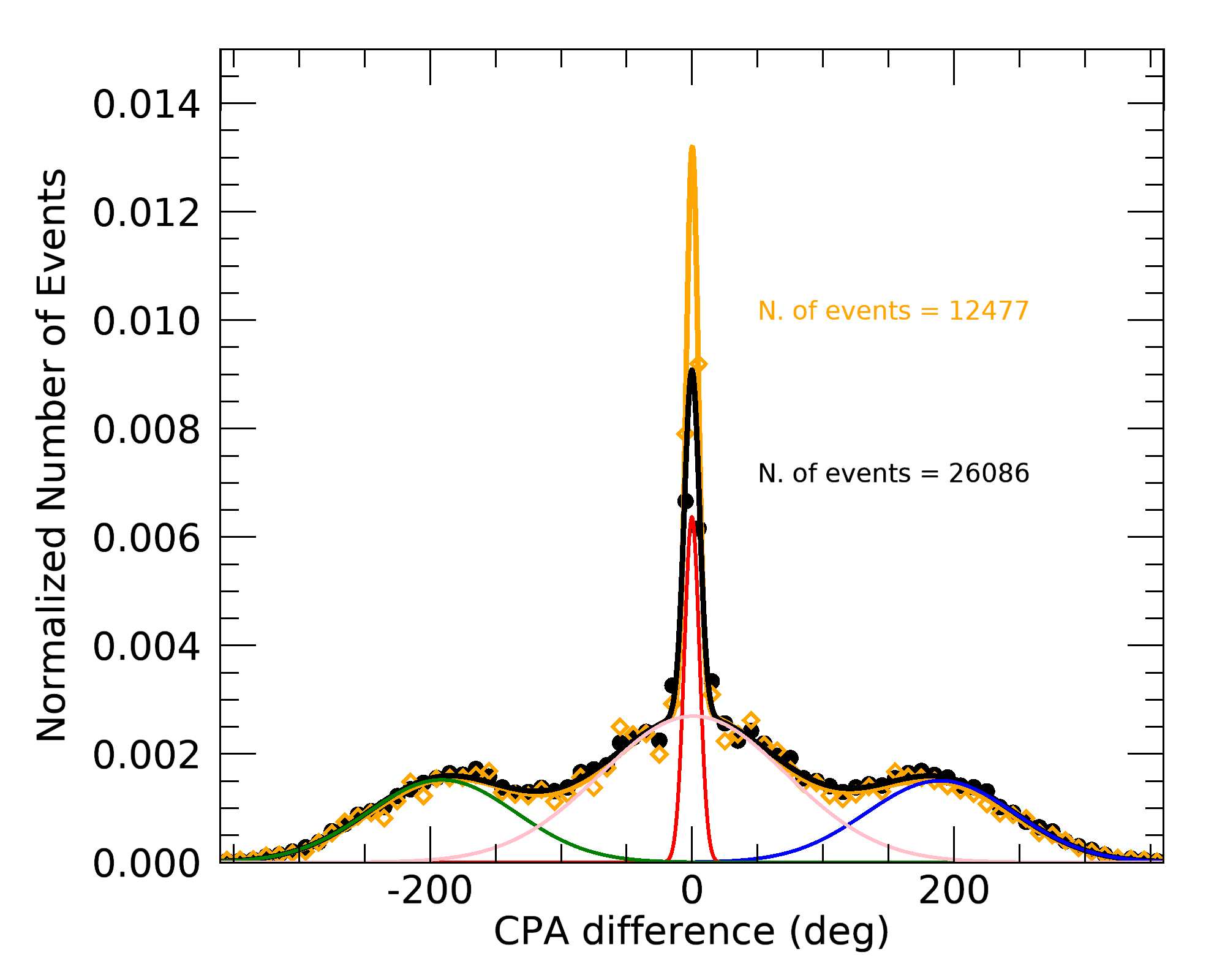}
\caption{Distribution of the difference of the position angle between any two consecutive CMEs with a bin size of $10^\circ$. The green, pink, red, and blue continuous lines represent the Gaussian distributions  added up (black curve) to fit the observations (black circles). The orange diamonds represent the observed distribution but for narrow (W $< 50^\circ$) CMEs.}
\label{fig:diff-cpa}
\end{figure}

The normalized distribution of the observed $PA_{diff}$  between consecutive CMEs is marked with black circles  in Figure \ref{fig:diff-cpa}. Over-plotted are the Gaussian PDFs,
\begin{equation}
  f(x)=\frac{1}{\sqrt{2\pi}\sigma} \exp\left(- \frac{(x-\mu)^2}{2 \sigma^2}\right), \label{eq:normal}
  \end{equation}
  which fit the data. We note that the three wide Gaussian distributions are centered at $\mu_{1,3} = \sim \pm 180^\circ$ (green and blue curves), and $\mu_2 \sim 0^\circ$ (pink curve)  representing
 the $PA_{diff}$ of randomly distributed events. What is unexpected for a random distribution of PAs is the narrow peak at the center of the distribution (around $0^\circ$), which can also be fit by a Gaussian (but narrower) distribution  shown by the red line in Figure \ref{fig:diff-cpa}. This large peak suggests that an important number of consecutive CMEs have less than $10^\circ$ of PA difference, meaning that these  CMEs are produced in a very close  source region  (even assuming that half of the events in the central peak of Figure \ref{fig:diff-cpa}
were produced on opposite sides of the Sun, this peak is $\sim 3$ times larger than the others
).
 It is worth noting that we have conserved the PA reference point (north pole) and the sign of the $PA_{diff}$  to facilitate the fitting process and to make the symmetries of the PA difference distribution clearer. Of course, the real angular distance ranges from  $0^\circ$ to $180^\circ$.
We are aware of the difficulty of using coronographs to give information about the real position, in the low corona, of the source region of CMEs. As the observations are projected in the plane of the sky, it is difficult to characterize the actual position of the CME source region. Nevertheless, the PA is a  good indicator of the radial direction of the CME, which in turn, extrapolated backwards to the solar surface, indicates the CME source region \citep[see also][]{2003ApJ...598L..63G,2008ApJ...688..647L,2018SoPh..293...60M}.

To diminish the effect of the projection, we performed the same
analysis using narrow (width $<50^\circ$) CMEs only
(we note that wherever we restrict the number of events,  the imposed condition has to be fulfilled by the two events of the leading-trailer pair of CMEs, if one of them does not meet the condition, the pair is discarded for that particular part of the analysis).
Obviously, the number of events is lower  (12477), but the statistical characteristics do not change, as shown by the observed  $PA_{diff}$ distribution of narrow CMEs plotted with diamonds and the associated Gaussian distributions (orange color) in Figure \ref{fig:diff-cpa}. In fact, the inference of a close source region of consecutive CMEs is more accentuated when we take into account only the narrow CMEs, as suggested by the larger height of the central peak.
These results suggest that the subset of CMEs with small $PA_{diff}$ may be spatially-related, in concordance with the scenario where new emerging flux frequently occurs in an active region which has already emerged \citep{1985SoPh...97...51L} or in its vicinity forming nests or clusters of flux emergence known as active longitudes \citep[see][and references therein]{vanDriel-Gesztelyi2015}.

\section{Speed difference of consecutive CMEs}
\label{sec:speed}

\begin{figure}
  \includegraphics[width=\columnwidth]{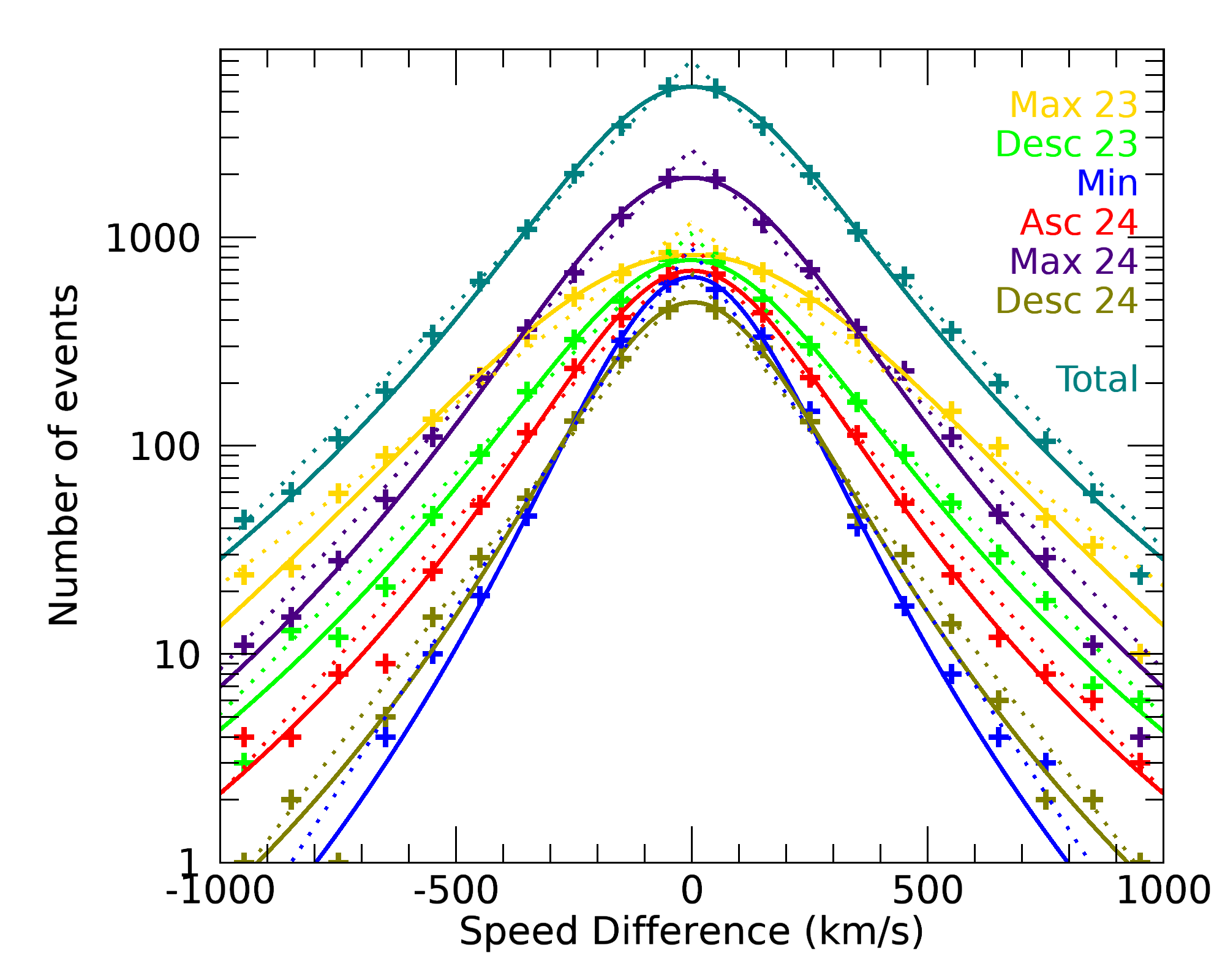}
  \caption{Observed (circles) and fit generalized student T- (continuous lines) and Laplace (dotted lines) distributions of the $V_{diff}$ during different phases of the solar cycle and the entire period of study (teal color).
  }
\label{fig:vdiff}
\end{figure}

Taking into account the short time difference  ($ WT \le 10$ hrs) between a vast majority of consecutive CMEs (note that after 10 hrs, a CME with speed of 500 km s$^{-1}$ has traveled  25 $R_\odot,$ and therefore remains inside the LASCO C3 field of view) and the close spatial relation between the source region  ($PA_{diff} \le 30^\circ$) of a group of them, a natural question  arises: Is there any relationship between the speed of the leading ($V_{lead}$) and trailing ($V_{trail}$ ) CMEs?
To explore this possibility, we tested several distributions and found that the  $V_{diff}= V_{trail} - V_{lead}$ best follows a generalized student T-distribution of the form:
\begin{equation}
f(x) = \frac{1}{\sqrt{\nu} 
\sigma  \textrm{Beta} (\frac{\nu}{2},\frac{1}{2})} \left[ \frac{\nu}{\nu +
    \left(  \frac{x-\mu}{\sigma}\right)^2  }\right]^{\frac{1+\nu.}{2}}
\end{equation} \label{eq:tstudent}
Figure \ref{fig:vdiff}
shows the observed (open circles) and the fitted T-distributions (continuous lines) of the $V_{diff}$ during the whole period (teal color) and during the individual periods of the solar cycle, plotted with different colors as marked in the Figure. The  corresponding
parameters are shown in Table \ref{table:student}.
It is important to note that the student T-distribution is also a heavy-tailed distribution. In this case, the proper equivalent for an exponential distribution is given by the Laplace distribution \citep[often known as the double-exponential distribution, ][]{9780470390634} of the form:

\begin{equation}
  f(x)= \begin{cases}
    \frac{1}{2\beta}\exp\left(\frac{-(x-\mu)}{\beta}\right)  & \text{if }  x \geq 0  \\
    \frac{1}{2\beta}\exp\left(\frac{-(\mu-x)}{\beta}\right)  & \text{if }  x < 0.
  \end{cases}
\label{eq:laplace}
\end{equation}

The BIC shows that the student T-distribution better fits the observed $V_{diff}$ than the double exponential distribution (the parameters of both distributions as well as the $\Delta$ BIC = BIC$_{Student}$ - BIC$_{Laplace}$ are shown in Table \ref{table:student}) and allows us to conclude that $V_{diff}$ distribution is heavy-tailed, and therefore, the associated CMEs may have long-range dependence.

 \begin{table*}
    \centering
  \caption{Parameters of the $V_{diff}$ distributions}\label{table:student}
  \begin{tabular}{l|ccc|ccc}
\hline
      & \multicolumn{3}{c}{Generalized T-Student} &
                                             \multicolumn{3}{c}{Laplace} \\ 
 Phase & $\mu_S$ & $\sigma$ & $\nu$ &  $\mu_L$ &  $\beta$ &
                                                                       $\Delta$ BIC \\
    \hline
Asc 23 & 7.18 & 195.90 & 3.41 & 5.00 & 203.66 & -4.40\\
Max 23 & -0.04 & 271.86 & 5.82 & -2.00 & 249.77 & -129.20\\
Desc 23 & -2.62 & 189.33 & 4.01 & -3.00 & 187.78 & -32.80\\
Min & -0.44 & 133.01 & 5.31 & -3.00 & 124.46 & -53.10\\
Asc 24 & -0.76 & 165.08 & 3.96 & 2.00 & 164.79 & -41.30\\
Max 24 & -1.73 & 180.24 & 4.44 & -2.00 & 173.72 & -23.00\\
Desc 24 & 1.28 & 149.26 & 4.50 & 2.00 & 143.99 & -22.80\\
Total & -1.13 & 186.15 & 3.91 & -1.00 & 185.37 & -62.00\\
\hline
  \end{tabular}
%  \label{table:student}
\end{table*}

\section{Long-range dependence}
\label{sec:dfa}

The  heavy-tailed distributions, such as the WT and the speed difference of consecutive CMEs, suggest that the underlying process may not be associated with a simple Poisson process, instead, some long-range dependence between the events may be present. In this section, we apply a test, often used to determine the presence of long-range dependence, on the CME variables.
Historically, the long-range dependence has been associated with time series where correlation decays slowly (decay exponentially in terms of the lag) or with a spectral density pole at zero frequency (i.e., at  the origin). Although, to apply these criteria, the time series under consideration must be stationary, and unfortunately in our case, this condition  is not fulfilled.
To overcome this restriction, the detrended fluctuation analysis (DFA) was proposed by
\citet{PhysRevE.49.1685} to study the long-range dependence of nucleoids of DNA chains. Basically, the goal is the determination of the scaling exponent $\alpha_s$  of the (log-log) dependence between the  detrended fluctuation function $F(s)$ and the size $s$ of the window where $F(s)$ was computed. The computational details can be found in \citet{2001PhyA..295..441K}.

\begin{figure}
 \includegraphics[width=\columnwidth]{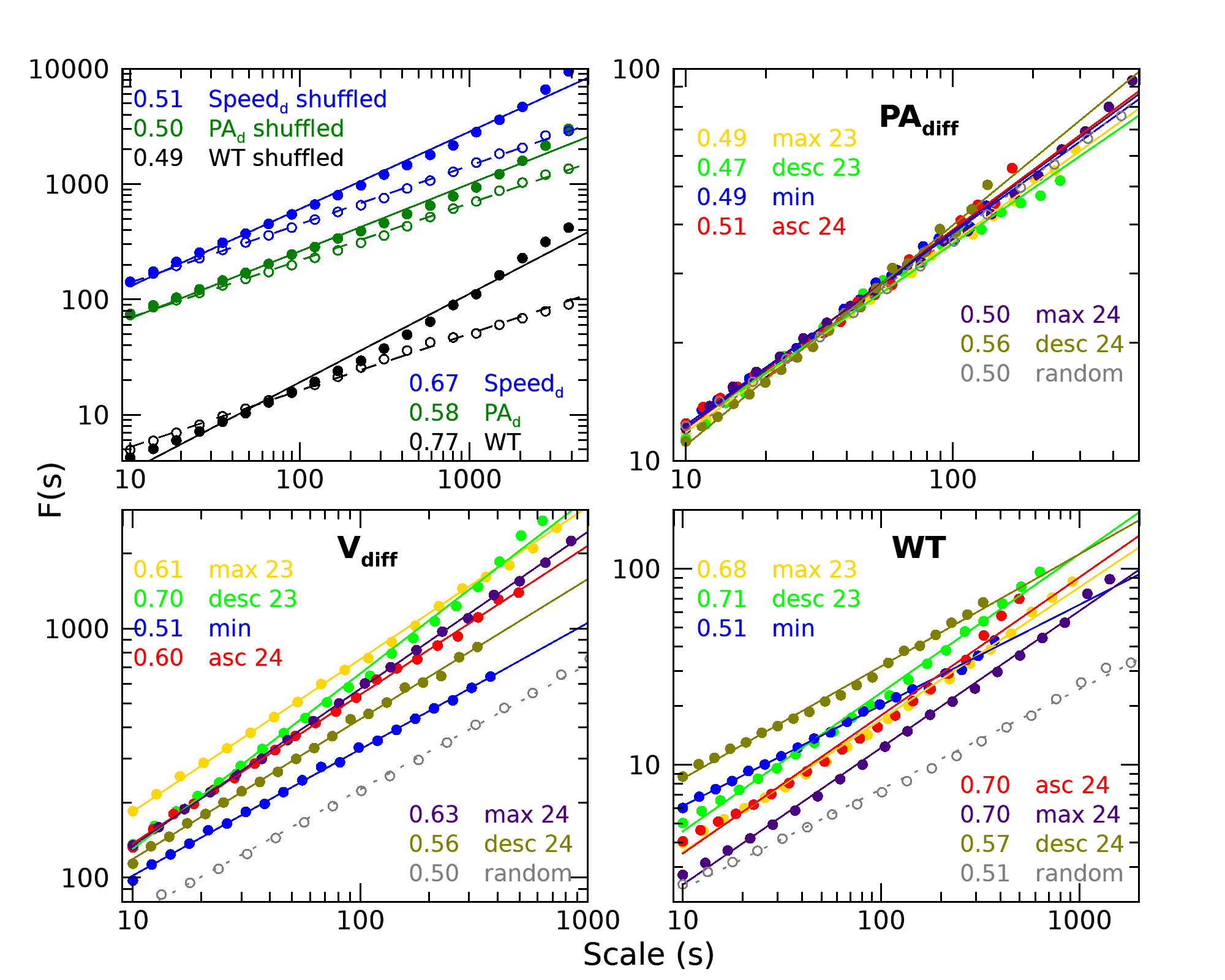}
 \caption{ Detrended fluctuation  as a function of the window size or scale (s) for the entire time period (upper left) of the WT (black), $V_{diff}$ (blue) and $PA_{diff}$ (green). The filled circles and continuous lines correspond to the actual data, whereas the empty circles and dashed lines correspond to the same, but randomly re-arranged (shuffled), data.
The same analysis but for each phase of the cycle applied to the $PA_{diff}$ (upper right), $V_{diff}$ (lower left) and WT (lower right). The $\alpha_s$ value during each phase of the cycle is marked with the corresponding color. Open circles and dashed lines correspond to the DFA applied to a random vector created for comparison purposes.}
%\end{subfigure}\par\medskip
\label{fig:dfa_tot}
\end{figure}

We applied the DFA to the WT, $PA_{diff,}$ and the speed
difference between the trailing and leading CME  ($V_{diff}$) % =V_{trail} - V_{lead}$) 
over the entire time range of study and the resulting DFA are shown in Figure \ref{fig:dfa_tot}, in black, blue, and green, respectively. Valuable information about the range dependence of the data is given by the scaling exponent. For a  random process such as Brownian motion,  $\alpha_s = 0.5$. This is the case of the $PA_{diff}$ and the randomly re-arranged WT and $V_{diff}$. The former is in accordance with its (sum of) Gaussian distribution.

On the other hand, $\alpha_s > 0.5$ implies long-range dependence. In this case, the $\alpha_s$ is 0.67 for the speed difference and 0.77 for the WT. To ensure  these results, we randomly shifted the values of those series and applied the DFA analysis, the results are plotted with open circles and dashed lines in Figure \ref{fig:dfa_tot} and for all cases $\alpha_s \sim 0.5$.

Similar to the probability distributions of the WT, $PA_{diff,}$  and $V_{diff}$, the scaling exponent changes with the phase of the solar cycle.
These changes are relatively small, for $PA_{diff}$, $\alpha_s \approx 0.5$ during all phases (upper-right panel of Figure \ref{fig:dfa_tot}). Similarly to during the minimum activity period, $\alpha_s \approx 0.5$ for both $V_{diff}$ and WT  (bottom-left and right panels, respectively). Conversely, the $\alpha_s$ reaches larger values during the maximum and descending phases of the solar cycle.

In summary, the DFA shows that there is long-range dependence on both the WT and $V_{diff}$  differences between successive CMEs. This is  evident during the maximum and declining phase of the solar cycle.

\section{Trailing and leading CME speed relationship}
\label{sec:speed_rel}

As suggested by the precedent analysis, consecutive CMEs have some dependence. In this section, we explore the possibility of finding a relationship between the speed of these events. In particular, the possibility that the speed of a  CME may be affected by a previous event.  For instance,
it may be surmised that the leading CME changes the ambient medium where the trailing CME is going to  travel, in such a way that the leading CME has an influence on the dynamics of the trailing CME as predicted by drag force models \citep{2004SoPh..221..135C,2009A&A...498..885B,2010A&A...512A..43V}.
At this point, it is important to recall that we do not apply any restriction to select the leading-trailing CME pairs, so that a leading can be a trailing of a previous event, and these are treated as an independent pair.
In order to investigate the possible statistical differences between the speed of the leading and trailing CMEs, we constructed subsets of spatial and temporally related pairs of CMEs with the following restrictions:
   i) A group with no restriction, meaning all events, hereinafter called $All_{CMEs}$.
  ii)  A group of  spatially-related consecutive CMEs where the PA difference ($PA_{diff}$) between the leading and trailing CMEs must be less than $30^\circ$.
  iii)  For both groups, the temporal association was established creating subsets of time windows where the WT is less than or equal to two to 25 hours.

\begin{figure}
  \includegraphics[width=\columnwidth]{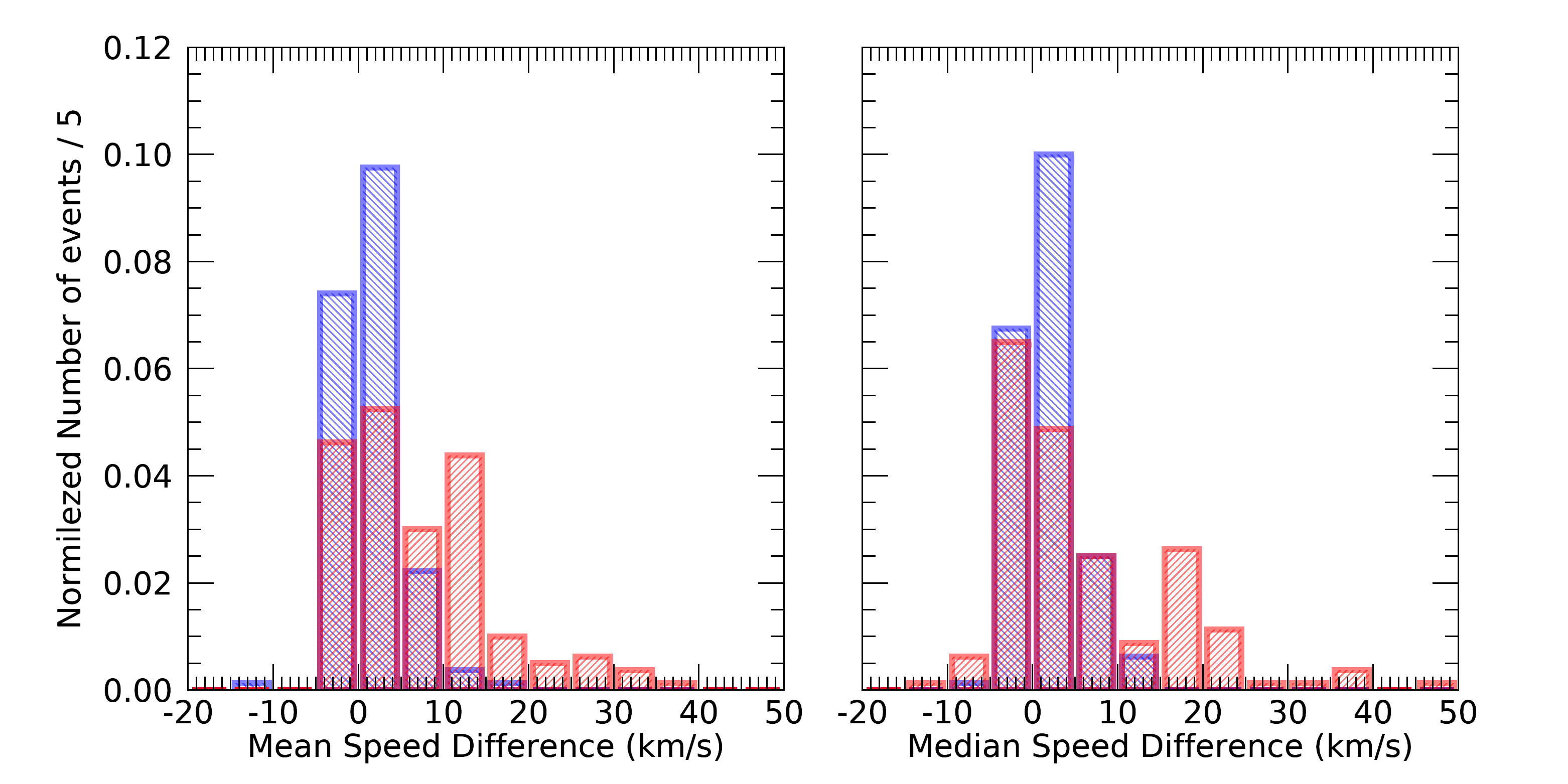}
\caption{Differences of trailing minus leading CME mean (left) and median (right) speed of  $All_{CMEs}$ (blue) and  spatially-related (red) subsets for all the considered WTs.}
\label{fig:vdif_hist}
\end{figure}

Then, we computed the mean speed value ($\overline{V}$) of each subset of leading and trailing CMEs and found a small but persistent difference between the $\overline{V}$ of the trailing and leading CMEs.
This   is shown  in the left panel of Figure \ref{fig:vdif_hist} where we plotted the (blue) histogram of the difference ($\overline{V_{diff}}$) between  trailing minus the leading CME means  (i. e., $\overline{V_{diff}} = \overline{V_{trail}} - \overline{V_{lead}}$) of the  $All_{CMEs}$ subsets. The main part of distribution is inside the -5 to 10 km s$^{-1}$ range, but with an asymmetrical component that extends towards  positive differences up to $\sim 17$ km s$^{-1}$. The asymmetry is evident for the   spatially-related CMEs  groups (red histogram),  where the difference attains higher values. \citet{2004JGRA..10912105G} found larger differences for the case of CMEs associated with solar energetic particles (SEP), although these were CMEs from the same active region.

\begin{figure}
  \includegraphics[width=\columnwidth]{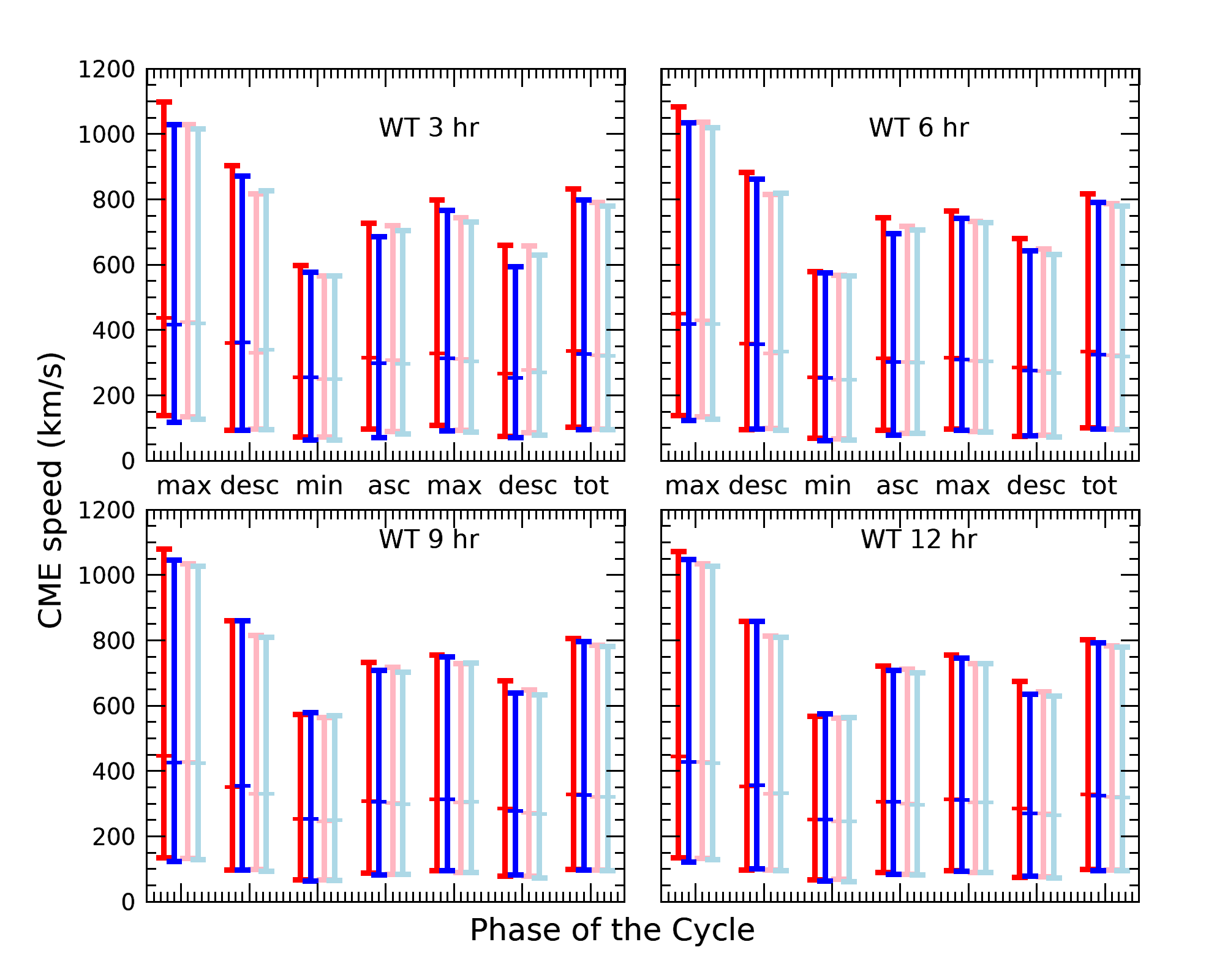}
\caption{Quartiles of speed distributions of 
the leading (light-blue) and  trailing (pink) $All_{CMEs}$ consecutive  pairs, as well as  spatially-related leading (blue) and trailing (red) consecutive CME pairs, as a function of the phase of the solar cycle for four different WTs (as marked in each panel).
}
\label{fig:quantil-vel}
\end{figure}

The CME speed is well-fit by a lognormal distribution \citep{2006JGRA..111.6107L}, which is also a heavy-tailed, and therefore the standard statistical tools, such as sample mean and sample standard deviation, are highly unstable \citep{pisarenko2010}.
In this case, the standard deviations of the speed distributions are  $\sim 200$ km s$^{-1,}$ and the standard errors of the mean are of the order of the differences. Then, a value of a few km s$^{-1}$ used to characterize the differences of the $\overline{V}$ has low statistical significance if  we use the mean  to characterize the speed distributions. Therefore, we have to use other parameters to quantify the observed differences.
The quartiles (particularly the second quartile or median) are better suited to characterize skewed distributions. So, we computed the quartiles of all subsets: for example, Figure \ref{fig:quantil-vel} shows the quartiles\footnote{This presentation is similar to the ``box-plot'', an often-used graphical method to illustrate the differences between distributions  where the quartiles (and the minimum and maximum values or ``whiskers'') of the  distributions that one wants to compare, are plotted side by side. The difference is that in this case, we do not plot the whiskers.} of the speed distributions as a function of the phase of the cycle for four different  WT subsets. Blue and red lines correspond to the trailing and leading  spatially-related CMEs; whereas pink and light-blue correspond to the  trailing and leading $All_{CMEs}$ groups, respectively.
As expected, the solar cycle variations are clear in the quartile behavior (Figure  \ref{fig:quantil-vel}), showing  higher values during the maxima, and the lowest values are reached during the minimum activity period. We note that the quartiles  clearly show the 
differences between the distributions of the leading and trailing CMEs.

For comparison (with the $\overline{V_{diff}}$), the right panel of Figure \ref{fig:vdif_hist} shows the differences between the medians of the trailing minus the leading speed distributions  (i. e., $\widetilde{V_{diff}} = \widetilde{V_{trail}} - \widetilde{V_{lead}}$). In this case, the maximum of the $\widetilde{V_{diff}}$ for the  spatially-related CMEs is shifted towards negative values, 
but the high asymmetry is conserved and attains higher values than the $\overline{V_{diff}}$.
The most important fact shown by both $\overline{V_{diff}}$ and $\widetilde{V_{diff}}$ is the asymmetry towards positive values, implying that the $V_{trail}$ tends to be higher than the $V_{lead}$ for consecutive CMEs.
In order to quantify these differences and their statistical significance,
we performed a sign test with the null hypothesis
 $H_0$:  
$ \widetilde{V_{trail}} - \widetilde{V_{lead}} \le 0$. It is unlikely  that $H_0$ is true, and therefore, the alternative hypothesis, in this case $H_a$: $ \widetilde{V_{trail}} - \widetilde{V_{lead}} > 0$, meaning $\widetilde{V_{trail}} > \widetilde{V_{lead}}$ is likely to be true.

\begin{figure}
  \includegraphics[width=\columnwidth]{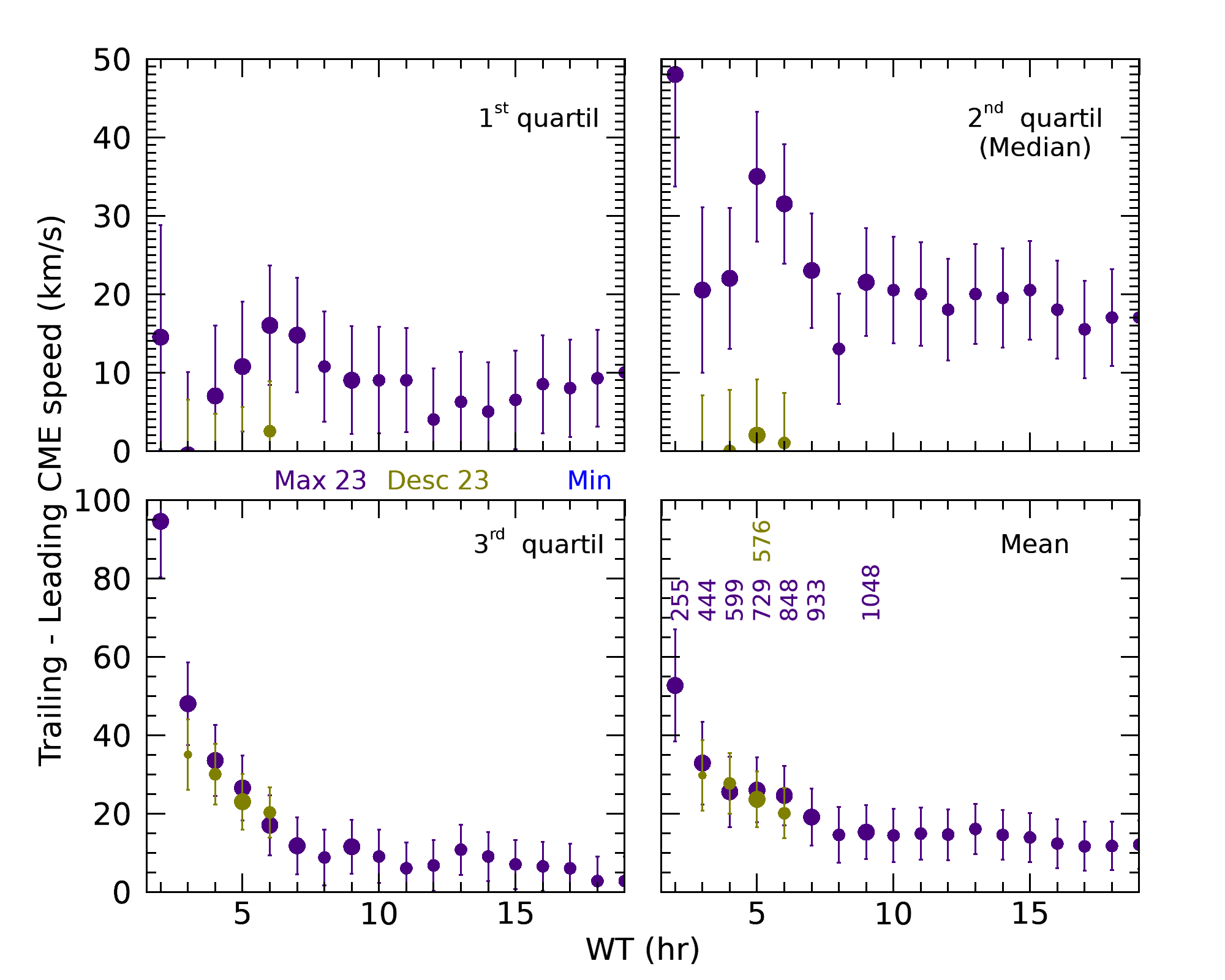}
  \caption{Difference between location parameters of the $V_{trail} -  V_{lead}$ distributions as a function of the WT. The three quartiles and the mean values during the minimum and two phases of solar cycle 23 are marked by different colors.
The size of the circles depends on the probability given by the sign test as: probability  $\le$  0.15 (small), 0.1 (medium), and 0.05 (large), i. e., when the null hypothesis ($H_0 : \widetilde{V_{trail}} - \widetilde{V_{lead}} \le 0$) is rejected at the 15, 10, and 5 \% level based in the sign test. The events considered in each subset with  probability $\le$  0.05 are also shown in the lower right panel.}
\label{fig:vdif_23}
\end{figure}

\begin{figure}
  \includegraphics[width=\columnwidth]{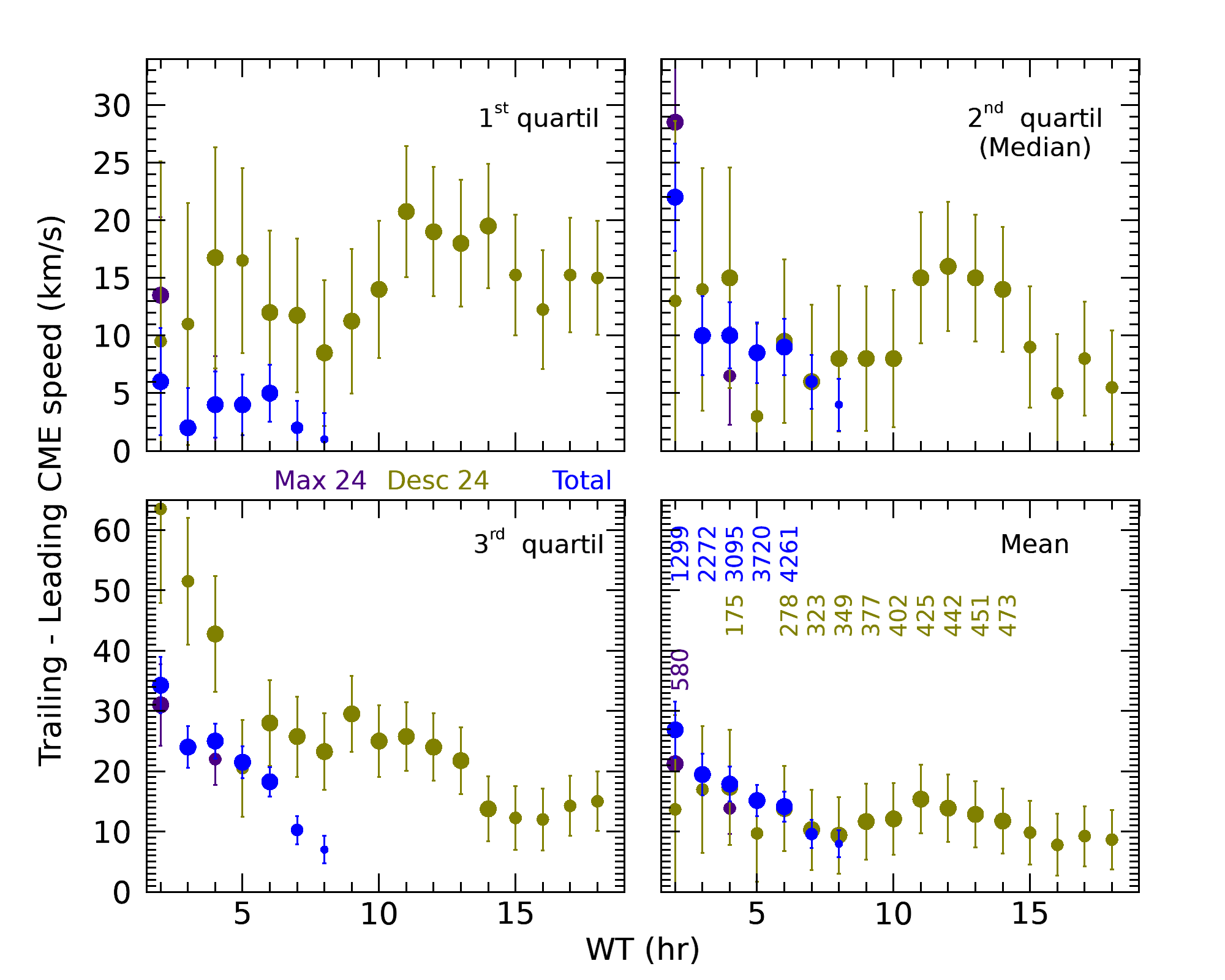}
  \caption{Same as Figure \ref{fig:vdif_23} but for solar cycle 24 and for the entire time range considered in this work (blue color).}
 \label{fig:vdif_24}
\end{figure}

The  $\widetilde{V_{diff}}$ and $\overline{V_{diff}}$ (upper- and bottom-right panels) as well as the  differences of the lower and upper quartiles (upper and bottom-left panels) are shown in Figures \ref{fig:vdif_23}  and \ref{fig:vdif_24} for cycles 23 and 24, respectively.
For the sake of clarity, we selected  the subsets 
whose median sign-test probability is lower than 0.15, 0.1, and 0.05,
marked with small, medium, and large symbols, respectively. Moreover, the number of events in each subset with probabilities lower than 0.05 are shown in the lower-right panel. The error bars in these Figures correspond to the median absolute deviation divided by the square root of the number of elements of each subset. 
Here, it is clear that the positive $V_{diff}$ is statistically significative,  and that there are excesses of up to $\sim 100$  km s$^{-1}$ for $WT \le 2$ hrs during the ascending and maximum phases of cycle 23, and a more modest excess of $\sim 20$  km s$^{-1}$ during the decreasing phase of cycle 24.
These $V_{trail}$  excesses decrease rapidly when the WT increases reaching  $\sim 20$  km s$^{-1}$ for cycle 23, and $\sim  0$  km s$^{-1}$ for cycle 24 when $WT < 10$ hrs.

The   spatially-related CMEs (marked with circles in Figures \ref{fig:vdif_23} and \ref{fig:vdif_24}) show the largest speed differences.
Overall, the  behavior of  the $V_{diff}$
is clearly seen in the bottom panels of Figures \ref{fig:vdif_23} and \ref{fig:vdif_24}, where the upper,  meaning the high velocity quartile (left), and the $\overline{V_{diff}}$ (right) differences, are shown. The speed difference has higher values for WTs $\le 2$ hrs and decreases in a quasi-exponential way up to WTs $\sim 7 - 8$ hrs. Then, it slowly decreases towards zero.
Finally, two important facts that maintain the CME speed bounded: (i) the speed difference reaches high values when we consider spatially-related CMEs, whereas the values remain low for all CMEs; and (ii) the speed difference tends to zero for large WTs during the minimum activity period.

\section{Discussion}
\label{sec:disc}

The statistical characterization of the properties of
consecutive CMEs is important because it sheds light on the generation processes and early stages of the CME phenomena.
In particular, the WT statistical distribution has been addressed by several authors \citep[see][and references therein]{2014ApJ...781L...1T}. Although, as far as we know, the associated PA and speed differences between successive CMEs have not been explored until now. A major subject of study is the nature of the CME initiation process: Is this a ``pure'' stochastic process? Or is there some kind of relationship between consecutive events? This possible relationship is called ``memory'' of the system or ``long-term dependence,'' and there are different tools to investigate if  the events of a given system  have this dependence. If the process and its time series  are stationary, the most common approaches used to look for  the existence of long-range dependence are the auto-correlation function and the power spectrum behavior at the origin. The CME process is not stationary, so we approached the problem through the shape of the statistical distribution tail and the DFA \citep[see][for extensive reviews of this subject]{pisarenko2010,Beran2013}. In this way,  a ``pure'' stochastic process has an exponential distribution and is considered as a memory-less system. This is the case  for the waiting times of a Poisson process \citep{LEE20051}, for example.
In contrast, processes where the distribution tail decreases more slowly than the exponential, the so called heavy-tailed distributions, such as the Pareto family and the generalized student T- distributions,  are often associated with long-range dependence or memory processes \citep{pisarenko2010,Beran2013}.

The main question in this statistical study can be formulated as: Is the CME process ``purely'' stochastic? Or does the CME process have a long-range dependence or ``memory'' in such a way that an event is influenced by the occurrence of past events?

To address this question, \citet{2003SoPh..214..361W} analyzed a set of CMEs observed by LASCO during the ascending phase and part of the maximum of cycle 23 (1996 - 2001).  The author fit the tail of the WT distribution  (WT $> 10$ hrs) by a power law function (with index $\gamma \approx -2.36$). The author also applied a Bayesian statistical analysis and  concluded that the observed WT distribution corresponds to a time-dependent Poisson process.
Performing a similar analysis but using LASCO observations in a restricted time range around the maximum of cycle 23 (1999 - 2001), \citet{2003ApJ...588.1176M} reached the same conclusion that the WT distribution of 3187 CMEs may be represented by a nonstationary Poisson process.

A  power law distribution, using only 113 CMEs observed by the Solar Maximum Mission, during the minimum phase of the cycle (1984 - 1989), was also found by  \citet{2006Ap&SS.302..213M}, with $\gamma = 1.41$. In this case, the authors associated the WT with a characteristic length $\Lambda$, (between two consecutive CMEs) as $\Lambda \propto V_a ~ WT$, where $V_a$ is the Alfv\'en speed (this implies causality), and  conclude that the WT power law distribution is consistent with their picture of characteristic length.

It is clear that a power law may be used to fit the tail of the WT distribution, but as mentioned earlier, 83\% of the total events have a WT under 10 hrs. Therefore, the tail after this WT time is not necessarily representative of the whole process (only the extreme events fall in the extended tail of the distribution). In fact, a power law distribution is heavy-tailed and may represent a scaling system, which, in turn, may have self-organized criticality or processes with long-range dependence or ``memory''  \citep{pisarenko2010,Beran2013}. For instance, \citet{Takahashi_2016} found a scaling relationship between CMEs and proton events. Also, \citet{2014ApJ...781L...1T} found that the WT distribution may be fit by a Weibull distribution (a heavy-tailed distribution), implying  some degree of correlation between CMEs in opposition to a purely stochastic process.
Our findings are in agreement with \citet{2009JGRA..11410105J} (who found that SEP events  appear to have some memory indicating that events are not completely random) and with the scaling relationship between CMEs and SEP found by \citet{Takahashi_2016}.

The possible correlation between consecutive CMEs is supported by the remarkable peak at short $PA_{diff}$, which shows that the source region of a large number of CMEs is located within close distance (Figure \ref{fig:diff-cpa}).
This result confirms the finding of \citet{2003ApJ...588.1176M} of an excess of events within an angular difference of $10^\circ$, these authors interpreted it as evidence of CMEs that occurred in the same active region. This result  is also in agreement with the hypothesis that many CMEs (and flares) are produced in the so-called active longitudes \citep[e.g.,][and references therein]{2017ApJ...838...18G}.

The WT distribution is telling us that, on average, the reconfiguration of the magnetic field after the launch of the leading CME, in order to replenish the free energy and launch the trailing CME, takes place in $\sim 3.7$ hrs during the maximum phase of cycle 24 and $\sim 8.5$ hrs during the minimum phase.
This is an important fact from the point of view of the CME energetics, in particular in the case where two or more consecutive CMEs originated in the same source region.  
It is worth to note that  the maximum fraction of free energy exhausted by one CME may be 25\% \citep{2005JGRA..110.9S15G,2012ApJ...759...71E,2013ApJ...765...37F}.

Finally, the speed  distributions  of all leading/trailing
CMEs show a small but systematic tendency to have higher  mean and median velocities for shorter WTs. The mean and median speed of the trailing CME are  a few km s$^{-1}$ higher than that of the leading CMEs, for WTs shorter than 7 hrs, after a WT of 10 hrs the differences remain constant and negligible. Nevertheless, these differences are higher (tens of km s$^{-1}$) when we only take into account   spatially-related CMEs (see Figure \ref{fig:vdif_hist}).

As the mean is not the best location indicator for highly skewed distributions, we also analyzed the quartiles (which are better suited to characterizing these distributions), the results are plotted in Figure \ref{fig:quantil-vel}. To assess the accuracy of the analysis,  we applied the sign test  to show that the differences between the medians of trailing and leading CMEs  are statistically significant. The phases of the solar cycle
where the  sign-test  probability  (with the null hypothesis $H_0 : \widetilde{V_{trail}} - \widetilde{V_{lead}} \le 0$) is less than 0.1 are shown in Figures \ref{fig:vdif_23} and \ref{fig:vdif_24}.

These observations may be explained by the 
drag force models. It is clear that the magnetic field provides the necessary energy to launch and first  acceleration of  CMEs. Although this energy does not play an important role in the CME dynamics after few solar radii, at least for fast CMEs (such as those that form the tail of the distributions).
Recently, \citet{2017SoPh..292..118S} investigated CME dynamics using coronographic observations, and they concluded that for fast CMEs, the Lorentz force has its maximum effect at heights lower than 2.5 $R_{\odot}$, and then, at heights of 3.5 - 4  $R_{\odot}$ becomes negligible compared with the drag force. In the case of slow CMEs, the Lorentz force may be significant at larger distances (up to 50 $R_{\odot}$, but slow CMEs do not contribute to the tail of the distributions). The drag force
$F_D = - \frac{1}{2} C_D \rho A (V_{CME} - V_{SW})^2$     \label{eq:dragforce}
is proportional to the drag coefficient ($C_D$), the transverse area of the CME ($A$), the speed difference between the CME ($V_{CME}$) and the solar wind ($V_{SW}$), and the ambient  density ($\rho$) \citep{2004SoPh..221..135C,2009A&A...498..885B,2010A&A...512A..43V}.

In a statistical analysis of a large number of events, the main values of $A$ and $V_{SW}$ are similar for both leading and trailing CMEs. On the other hand, the leading CME may  sweep the ambient medium, and the trailing CME encounters a less dense medium and therefore the drag force decreases, resulting in a faster CME. In this case, the shorter the WT, the higher the CME speed. This effect is clearly seen in Figures \ref{fig:vdif_23} and \ref{fig:vdif_24}.

\section{Conclusions}
\label{sec:conclusions}
We performed a statistical study of  27761  CMEs observed by SOHO/LASCO from the ascending phase of cycle 23 to the declining phase of cycle 24, looking for long-range or memory signatures in the genesis and the first stages of evolution of CMEs.
The main findings of this work are:
\begin{itemize}
  \item The time elapsed between consecutive CMEs or WT for 98\% (61\%) of all events is less than 25 (5) hrs.

\item We found that a Pareto Type IV probability density function
fits the observed WT distribution 
during the entire period (1996 - 2018), and during the different  phases of the solar cycle.

\item As expected, the WT varies with the solar cycle, with large mean values ($\sim 10$ hrs) and high rates of change over time at low activity phases, and short mean ($\sim 4$ hrs) and $\sim 0$ rates of change over time when the activity is high.
These mean values are consistent with recurrence times in large active regions reported before \citep[see Figure 3 in][]{2005JGRA..110.9S15G}.

\item The distribution of the PA difference of consecutive CMEs (as a proxy to the CME source  region) shows four components: three wide Gaussian distributions centered at $0^\circ$ and $\pm 180^\circ$ related to the spatially uncorrelated CMEs; the fourth is  a narrow Gaussian centered at $0^\circ$ related to CMEs with very close source regions.
  
\item The difference between the trailing and leading CME speed follows a generalized student T-distribution.

\item The heavy-tailed distributions as well as the DFA show that there is a long-range dependence or memory in  the WT and speed difference of consecutive CMEs.
  
\item The long-range dependence of the CME speed is confirmed by the strong statistical evidence that the speed of the trailing CMEs is  slightly higher than that of leading CMEs where the WTs are $< 10$ hrs.

\item The trailing or leading speed  difference is higher for the CMEs with close source-region association. 
\end{itemize}

These findings point towards the fact that leading CMEs modify (sweep up) the ambient medium, causing the trailing CMEs to encounter less opposition to their movement. This difference is clear for CMEs produced by individual active regions \citep[see Figure 9 of][]{2004JGRA..10912105G}.

In summary, by analyzing a large number of  consecutive CMEs, we find that their WT distribution follows a Pareto Type IV distribution, and the associated speed difference follows a generalized student T-distribution. These heavy-tailed distributions along the DFA suggest long-term dependence in the CME process. The position-angle  difference distribution shows a large Gaussian peak centered at zero, indicating a close spatial relationship between consecutive CMEs. Furthermore, there is a small but statistically significant difference between the speed of consecutive CMEs indicating that the trailing CME is  a few
km s$^{-1}$ faster than the leading CME when the WT is $< 10$ hrs. All these findings suggest a physical connection between a considerable population of  consecutive CMEs.

%%%%%%%%%%%%%%%%%%%%%%%%%%%%%%%%%%%%%%%%%%%%%%%%%

\begin{acknowledgements}

The CME catalog is generated and maintained at the CDAW Data Center by
NASA and The Catholic University of America in cooperation with the
Naval Research Laboratory. SOHO is a project of international
cooperation between ESA and NASA.  This work was partially supported
by CONACyT (179588) and UNAM-PAPIIT (IN111716-3). N. Gopalswamy is
supported by NASA's Heliophysics LWS and GI programs. We thank the
anonymous referee for his/her useful and constructive comments.
\end{acknowledgements}

\bibliographystyle{aa}
\bibliography{draft}

\end{document}